\documentclass[letter,prd,notitlepage,12pt,showpacs,showkeys]{revtex4-1}

\usepackage[]{graphicx}
\usepackage{amsmath}

\newcommand{\Ethr}{\epsilon_{\text{thr.}}}
\newcommand{\meanLnA}{\left<\ln{A}\right>}
\newcommand{\tsectheta}{\times\sec{\theta}}
\newcommand{\rhos}{\rho_{s,600}}
\newcommand{\rhom}{\rho_{\mu,300}}
\newcommand{\Rm}{R_{\text{M}}}
\newcommand{\Xm}{x_{\text{max}}}

\begin{document}

\title{Variations of the cosmic ray composition at energy above 0.1~EeV as observed by muon detectors of Yakutsk array}

\author{A. V. Glushkov}
\author{A. Sabourov}
\affiliation{Yu. G. Shafer Institute of cosmophysical research and aeronomy}
\address{677980, Lenin Ave. 31, Yakutsk, Russia}
\email{tema@ikfia.sbras.ru}

\begin{abstract}
  The lateral distribution of muons with $\sim 1.0\tsectheta$~GeV in extensive air showers within $\sim 10^{17} - 10^{19}$~eV energy region obtained during different observational periods from November 1987 to June 2013 has been analyzed. Experimental data have been compared to predictions of various hadron interaction models. The best agreement is observed with QGSJETII-04. Until 1996, the mass composition of cosmic rays with energy below $2 \times 10^{18}$~eV was significantly lighter than in later periods. 
\end{abstract}



\maketitle

\section{Introduction}

Ultra-high energy ($E \ge 10^{15}$~eV) cosmic rays (UHECR) are still remain a major scientific problem despite being studied worldwide by extensive air shower (EAS) arrays for good 50 years. Their mass composition is still not known exactly, and without this knowledge it is difficult to understand the character of nuclear interactions in this energy region. Muons with energy near $0.5 - 1.0$~GeV are very important component of EAS. They are poorly moderated in the atmosphere, are sensitive to the characteristics of nuclear interactions during development of a shower and to the chemical composition of cosmic rays (CR). Due to their yield and properties of lateral distribution they can be effectively registered with widely spaced ground arrays. Since 1978 Yakutsk EAS array has been continuously registering muons with the threshold energy $\Ethr \simeq 1.0\tsectheta$~GeV. During this period a large amount of experimental data has been accumulated. Analysis of this material~\citep{Astropart95, JETPl98, JETPl2000, YaF2000, YaF2005, JETP2006} has revealed that the development of showers with $E_0 \ge (3-5)\times 10^{18}$~eV differs significantly from those at lesser energies. It also allowed us to estimate the fraction of primary gamma-quanta in the total CR flux at energies above $10^{17}$~eV~\citep{YakutskINR}.

\begin{figure}
  \centering
  \includegraphics[width=0.95\textwidth, clip]{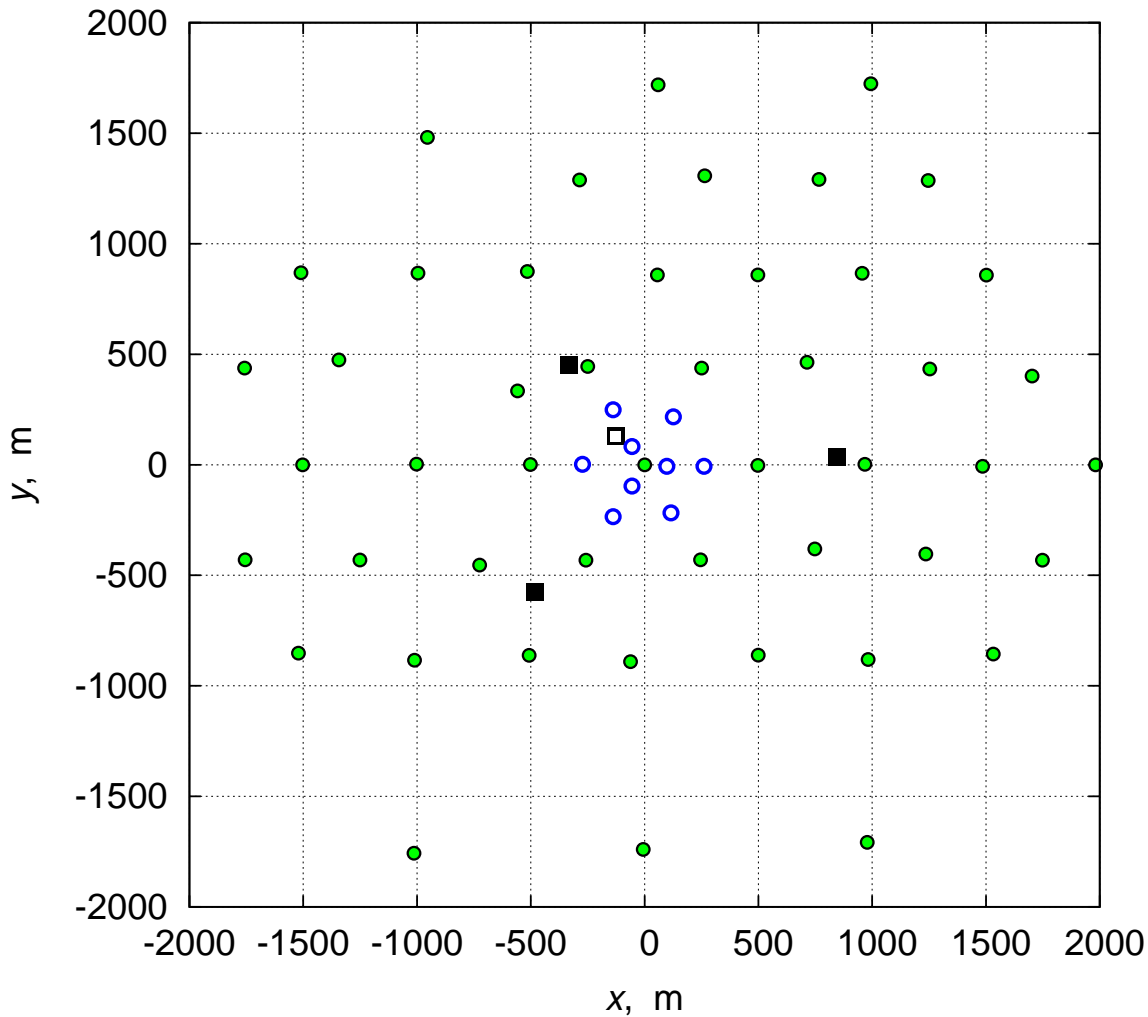}
  \caption{The layout of detectors location at Yakutsk array (since 1992). Green circles represent detectors of the main array ($2\times 2$~m$^2$); empty blue circles~--- additional scintillation detectors ($2$~m$^2$); filled black squares~--- underground muon detectors of $10\times 2$~m$^2$ area and $1.0\times\sec{\theta}$~GeV threshold; empty black square~--- underground muon detector of $27\times 2$~m$^2$ and $0.5\times\sec{\theta}$~GeV threshold.}
  \label{fig:1}
\end{figure}

Recently we have shown \citep{JETPl2012} that the muon fraction in the total number of charged particles in EASs with energies $E_0 \ge 10^{17}$~eV changes significantly over periods of time. Until 1996, it fluctuated around a single stable position and then increased significantly. This was accompanied by near simultaneous variations in the energy spectrum and in the global anisotropy of CR within energy range $(1-10)\times 10^{17}$~eV~\citep{JETPl2012,AstroLet2013}. After 1996, during the next 7 years, the integral intensity of CR at $E_0 = 10^{17}$~eV increased by $(45 \pm 5)$\,\% and then started declining. As for the phase of the first harmonic $\phi_1 = 119^{\circ} \pm 18^{\circ}$ and its amplitude $A_1 = 0.030 \pm 0.014$ sampled during 1983-1994, they changed to values $\phi_1 = 284^{\circ} \pm 13^{\circ}$ and $A_1 = 0.033 \pm 0.010$ during 1998-2010. In recent years, a tendency has been manifested towards the change of these values in the opposite direction. It seems like the aftermath of some gargantuan explosion which have contributed a significant portion of heavy nuclei to the background. It's still difficult to pinpoint exactly what kind of event in the Galaxy could led to such result. Here we need further studies involving temporal factor of the experimental data. In this work we present our estimations of the mean CR composition obtained from the analyzes of muon data with $\sim 1.0\tsectheta$~GeV threshold which have been gathered during different periods of their operation from 1987 to June 2013. The geometry of Yakutsk array including its muon detectors is presented on Fig.\ref{fig:1}. The teqhnique of their control and calibration is described in the~\citep{JETPl2012}. The data from the detector with $\Ethr \simeq 0.5\tsectheta$~GeV~\citep{JETPl2013} are currently being accumulated and will be analyzed later.

\section{Results}

\begin{figure}
  \centering
  \includegraphics[width=0.85\textwidth]{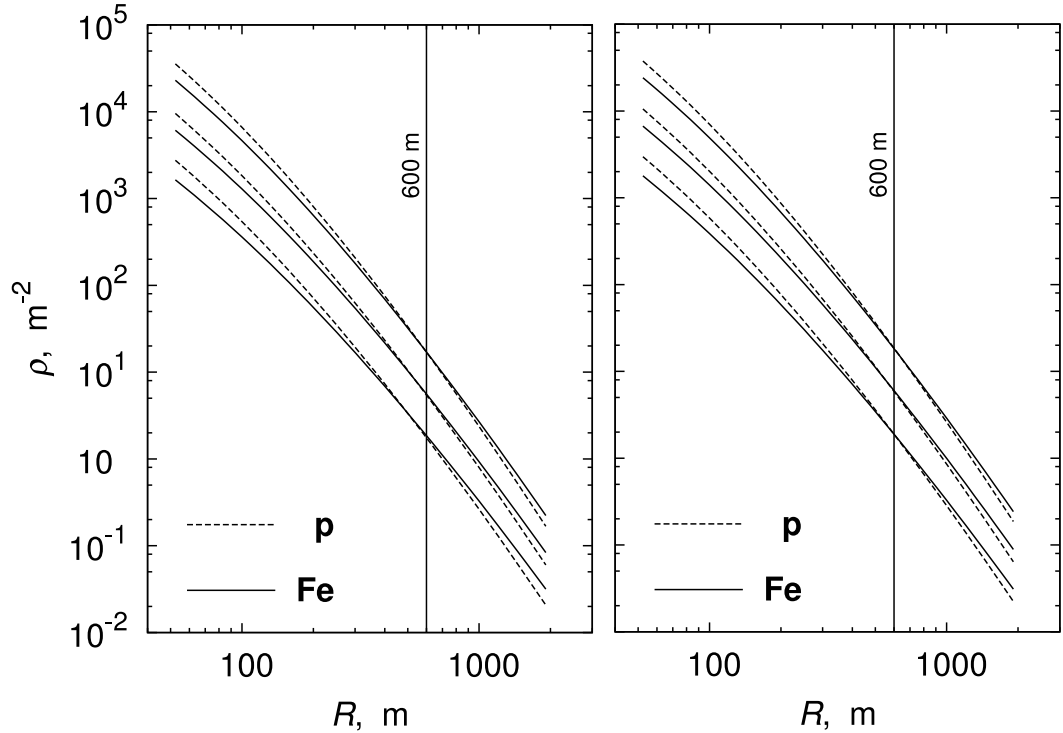}
  \caption{LDFs of charged particles in showers with energies $10^{18}, 3.16\times 10^{18}$ and $10^{19}$~eV at the cosine of zenith angle $\cos{\theta} = 0.9$ from primary protons and iron nuclei obtained within the frameworks of QGSJET01D~\cite{QGSJET} (left) and QGSJETII-04~\cite{QGSJETII} (right) models.}
  \label{fig:2}
\end{figure}

For this work we considered EASs with zenith angles $\theta \le 45^{\circ}$ and axes fallen within a $1$~km radius circle in the center of the array and with the precision of axis detection no less than~$20$~m. The energy of primary particles was determined from relations:
\begin{equation}
  E_0\text{, eV} = (4.8 \pm 1.6) \times 10^{17} \cdot \rhos(0^{\circ})^{1.0 \pm 0.02}~\text{,}
  \label{eq:1}
\end{equation}
\begin{equation}
  \rhos(0^{\circ})\text{, m}^{-2} = \rhos(\theta) \cdot \exp{\frac{(\sec{\theta - 1}) \cdot 1020}{\lambda_{\text{p}}}} \text{}~\text{,}
  \label{eq:2}
\end{equation}
\begin{equation}
  \lambda_{\text{p}}\text{, g/cm}^2 = (450 \pm 44) + (32 \pm 15) \cdot \log_{10}{\rhos(0^{\circ})}~\text{,}
  \label{eq:3}
\end{equation}
where $\rhos(\theta)$ is the density of all charged particles (electrons and muons) as measured by surface scintillation detectors at $R = 600$~m from a shower axis. The precision of $\rhos$ estimation in individual showers was no worse than 10\,\%. The relation~(\ref{eq:1}) unambiguously connects the $\rhos(0^{\circ})$ with $E_0$ at any given CR composition. It is due to the fact that at the distance $\sim 600$~m from the axis, the lateral distribution functions (LDF) of charged particles inter-cross each other, i.e. give the same value regardless of the type of a particle initiated a shower. It is demonstrated on Fig.\ref{fig:2}, where LDFs of charged particles are shown, as predicted by QGSJET01D~\cite{QGSJET} and QGSJETII-04~\citep{QGSJETII} models, in showers with $E_0 = 10^{18}-10^{19}$~eV at $\cos{\theta} = 0.9$ originated from protons (dashed line) and iron nuclei (solid line). Values for $\rhos(\theta)$ were derived from the modified Linsley approximation~\citep{Linsley62}:
\begin{equation}
  f_{s}(R,\theta) = \rhos(\theta) \cdot \frac{600}{R} \cdot
  \left(
    \frac{\Rm + 600}{\Rm + R}
  \right)^{b_s - 1}\text{,}
  \label{eq:4}
\end{equation}
where $\Rm$ is the Molier radius which depends on air temperature ($T, ^\circ$\,C) and pressure ($P$, mbarn):
\begin{equation}
  \Rm\text{, m} \simeq \frac{7.5 \times 10^4}{P} \cdot \frac{T}{273}~\text{.}
  \label{eq:5}
\end{equation}
The value for $\Rm$ is measured in each individual event (for Yakutsk $\left<T\right> \simeq -18^{\circ}$\,C, $\left<\Rm\right> \simeq 70$~m). In the expression (\ref{eq:4}), $b_s$ is the parameter defined in~\citep{Yakutsk76}:
\begin{equation}
  b_s = 1.38 + 2.16 \times \cos{\theta} + 0.15 \times \log_{10}{\rhos(\theta)}\text{.}
  \label{eq:6}
\end{equation}

\begin{figure}
  \centering
  \includegraphics[width=0.85\textwidth]{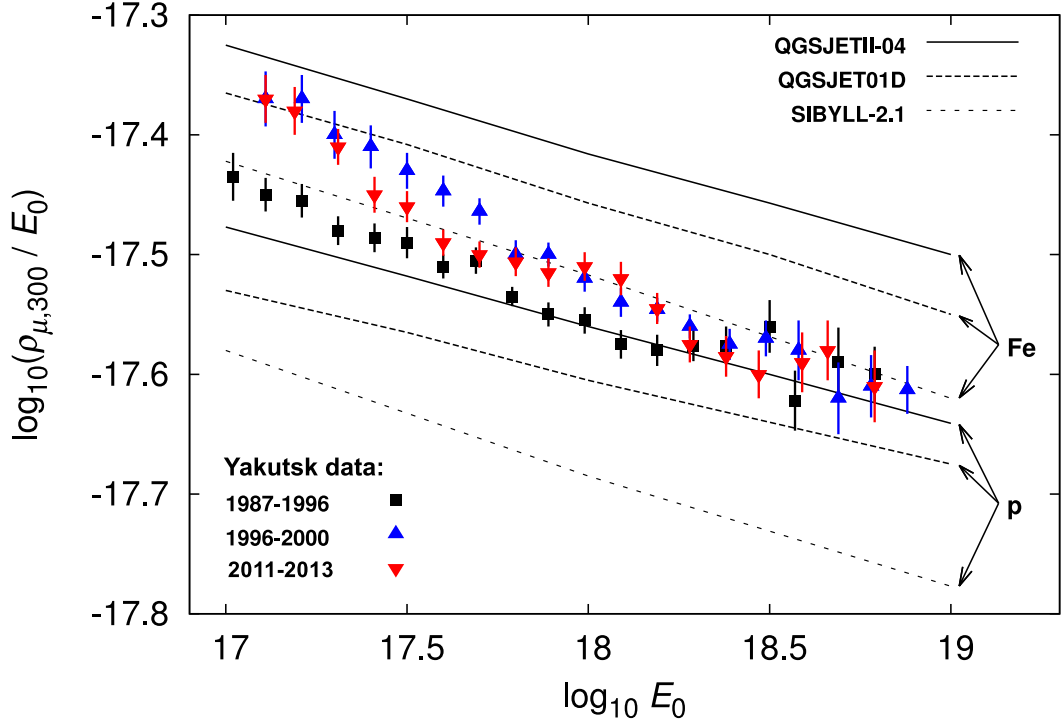}
  \caption{Densities of muons with $1.0\tsectheta$ threshold energy at 300~m from shower axis normalized to $E_0$ in showers with $\left<\cos{\theta}\right>=0.9$ and different primary energies.}
  \label{fig:3}
\end{figure}

On Fig.\ref{fig:3}, muon densities at $300$~m from shower axis $\log_{10}\left(\left<\rhom\right>/\left<E_0\right>\right)$ are shown for showers within groups with mean values $\left<E_{0}\right>$ and $\left<\cos{\theta}\right> = 0.9$. Normalization to primary energy gives a representation of muon data that is more descriptive and convenient for further analysis. Mean LDFs were obtained within energy bins with logarithmic step $h = \Delta\log_{10}{E_{0}} = 0.2$ which were subsequently shifted towards higher energies by the value $0.5\,h$. This procedure was performed in order to provide a detailed test of the agreement between the experiment and various hadron interaction models. Values of $\left<\rhom\right>$ were obtained from approximations of mean LDFs. When constructing an LDF, muon densities were multiplied by normalizing ratio $\left<E_0\right> / E_0$ and averaged over an energy cut in radial bins $\Delta\log_{10}{R} = 0.04$. Mean muon densities were determined from the expression
\begin{equation}
  \left<\rho_{\mu}(R_i)\right> =
  \frac{\sum_{n = 1}^{N_1} \rho_{\mu}(R_i)}{N_1 + N_0}\text{,}
  \label{eq:7}
\end{equation}
where $N_1$ and $N_0$ are the numbers of operated muon detectors at axis distances within the interval $(\log_{10}{R_i}, \log_{10}{R_i} + \Delta\log_{10}{R})$. The indexes denote whether a detectors had non-zero $(N_1)$ or zero $(N_0)$ readings during the registration of event. Zero readings are related to cases when a detector hasn't registered any muons while being in a wait state. Mean LDFs were approximated according to functions~\citep{YaF2000}:
\begin{equation}
  \rho_{\mu} = f_{\mu}(R,\theta) \cdot \left(1 + \frac{R}{2000}\right)^{-6.5}
  \label{eq:8}
\end{equation}
with well-known relation by Greisen~\cite{Greisen60}:
\begin{equation}
  f_{\mu}(R,\theta) = C_{\mu}\,N_{\mu}\,r^{-0.75} \cdot (1 + r)^{0.75 - b_{\mu}}\text{,}
  \label{eq:9}
\end{equation}
where $C_{\mu}$ is a normalization constant, $N_{\mu}$~--- full number of muons at the observational level ($1020$~g/cm$^2$ for Yakutsk), $r = R / 280$ and $b_{\mu}$ is a free parameter. The best fit values of $b_s$, $\rhos(\theta)$ in formula (\ref{eq:4}) and $b_{\mu}$, $\rho_{\mu,600}(\theta)$ in formula (\ref{eq:9}) were determined with the use of $\chi^2$ minimization. Error bars on Fig.\ref{fig:3}a include the entire combination originated from statistics of events and from averaging of the data. Lines represent expected values predicted by hadron interaction models QGSJETII-04 (solid), QGSJET01D~\cite{QGSJET} (dashed) and SIBYLL-2.1~\cite{SIBYLL} (dotted). Simulations were performed with the use of CORSIKA code~\cite{CORSIKA} (version 6.990 in the case of SIBYLL-2.1 and QGSJET01D and 7.3700 in the case of QGSJETII-04). 200 showers were simulated per each set of initial shower parameters (mass of primary particle, energy and zenith angle). To speed-up the computations, the thin-sampling algorithm was activated in the CORSIKA code with the parameters $E_i / E_0 \in [3.16 \times 10^{-6}, 10^{-5}]$ and $w_{\text{max}} \in [10^4, 3.16 \times 10^{6}]$ depending on the primary energy~\cite{CORSIKAUG}. The density was calculated directly from total number of particles arrived at a detector of given area.

It is seen that the experiment is not consistent with SIBYLL at neither given CR composition. This model predicts significantly lower muon yield. Other two models agree with our experiment much better and allow to estimate the mass composition of primary particles. To simplify, let us consider a two-component composition, consisting of protons and iron nuclei. In this case the relation
\begin{equation}
  \meanLnA = W_{\text{p}} \cdot \ln{1} + W_{\text{Fe}} \cdot \ln{56}
  \label{eq:10}
\end{equation}
gives weighting functions $W_{\text{p}} = 1 - W_{\text{Fe}}$ and $W_{\text{Fe}} = \meanLnA / \ln{56}$. Within the framework of this hypothesis we have:
\begin{equation}
  W_{\text{Fe}} = \frac{d_{\text{exp}} - d _{\text{p}}}
  {d_{\text{Fe}} - d_{\text{p}} }~\text{,}
  \label{eq:11}
\end{equation}
where $d = \log_{10}\left({\rhom} / E_0\right)$~--- are the values obtained in the experiment (exp) and in simulation.

\begin{figure}
  \centering
  \includegraphics[width=0.84\textwidth]{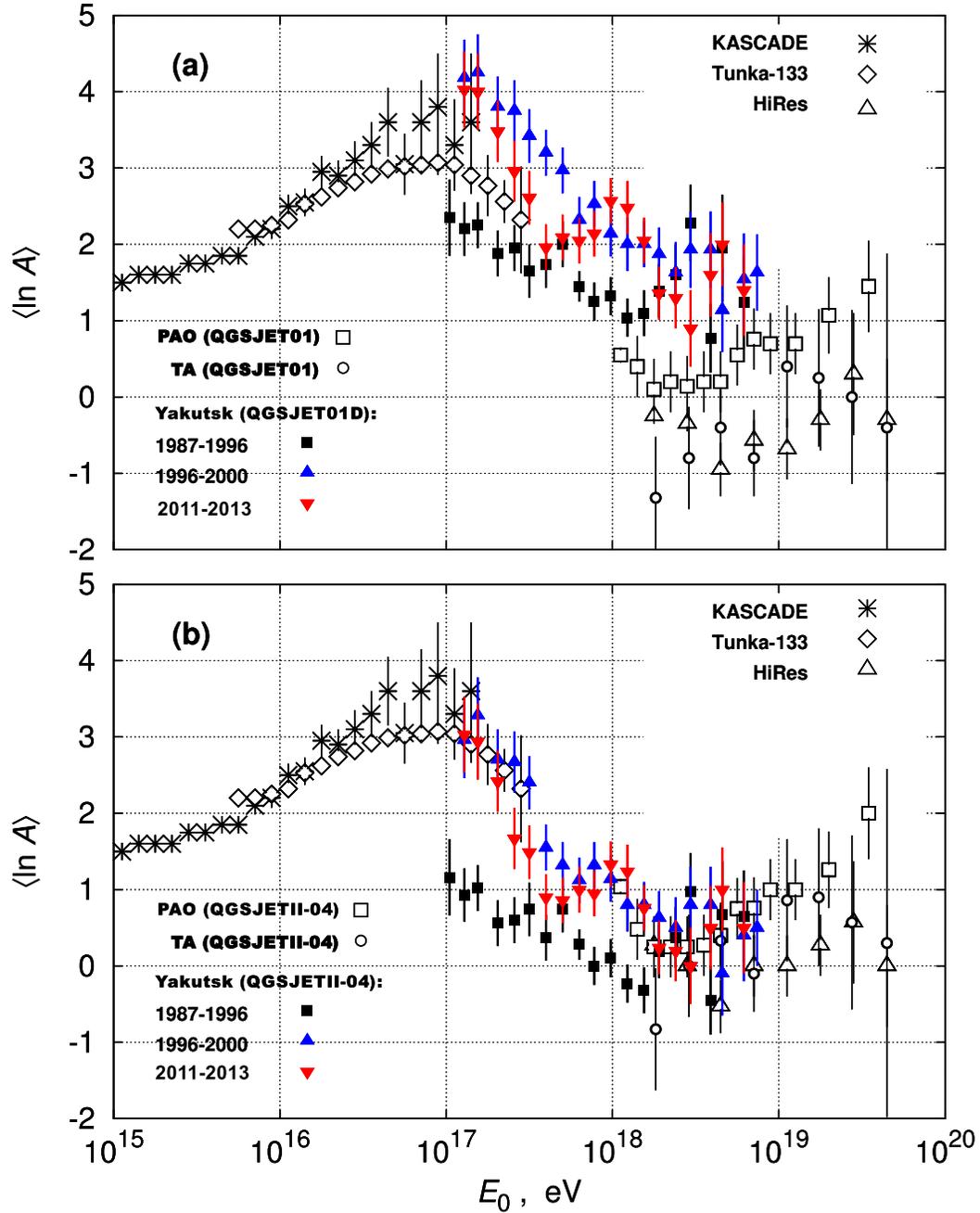}
  \caption{Mean atomic number of CR versus the energy of primary particles according to various experiments compared to our estimation based on $\left<\rhom\right>$ (for $1\tsectheta$~GeV threshold) for three observational periods within the frameworks of QGSJET01D {\bf(a)} and QGSJETII-04 {\bf (b)} models.}
  \label{fig:4}
\end{figure}

With black squares, blue and red triangles on Fig.\ref{fig:4} are shown energy dependencies of the CR mass composition obtained from the data of Yakutsk experiment during three observational periods according to predictions of QGSJET01D (Fig.\ref{fig:4}a) and QGSJETII-04 (Fig.\ref{fig:4}b) models. Black squares refer to the period 1987-1996 that directly preceded the above mentioned events described in~\citep{JETPl2012, AstroLet2013}. Blue triangles describe subsequent changes of the CR mass composition during 1996-2000 period and red triangles~--- its current state (2011-2013). One can see that after 1996 the mass composition of CR with energy up to $2\times 10^{18}$~eV became significantly heavier than before. Since then a tendency has been manifested towards its change in reverse. As for the region of $E_{0} \ge 2 \times 10^{18}$~eV, within the measurement errors, no significant changes were observed.

Stars on Fig.\ref{fig:4} represent the results of KASCADE obtained during the period from May 1998 to December 1999~\cite{Ulrich2001}. White diamonds represent the data of Tunka-133 experiment obtained from the Cherenkov light LDF during two winter seasons (2009-2011)~\cite{Tunka2011}. Other numbers were derived from experimentally measured values of $\Xm(E_{0})$ according to QGSJET01 and QGSJETII-04~\cite{TA2013} with the use of expression (\ref{eq:11}) with substitution $d = \Xm$. With white triangles are shown HiRes data related to the observational period from November 1999 to September 2001~\cite{HiRes2010}. White squares~--- results of PAO obtained between December 2004 and September 2010~\cite{PAO2011}, open circles~--- interpretation of the Telescope Array data.~\cite{TA2013}.

It is clearly seen that all estimations of the CR mass composition based on prediction of the QGSJETII-04 model show a better agreement with each other than those based on QGSJET01. The results presented on Fig.\ref{fig:4}b give evidence that since 1996 the CR composition in energy region $E_{0} \simeq (1 - 20) \times 10^{17}$~eV has been changing rapidly towards lighter nuclei with increasing primary energy. During 2009-2013 it decreased from $\meanLnA = 3.0 \pm 0.4$ at $E_{0} \simeq 10^{17}$~eV to the value $\meanLnA = 0.4 \pm 0.4$ at $E_{0} \simeq 2 \times 10^{18}$~eV. Earlier, according to our observations in 1987-1996, the composition used to be lighter ($\meanLnA = 1.0 \pm 0.4$ at $E_{0} \simeq 10^{17}$~eV). These numbers agree with the results presented in~\cite{JETPl2012, AstroLet2013} and a hypothesis of some gigantic explosion that added a significant fraction of heavy nuclei to the background.

\section{Conclusion}

The comparison between the muon data obtained at Yakutsk array and modern ultra-high energy interaction models has demonstrated once again the importance that this component presents for studying of extensive air shower development and CR mass composition. The results from Fig.\ref{fig:3} demonstrate a certain degree of agreement between the experiment and QGSJETII-04 and QGSJET01D models within the energy range $E_0 \simeq 10^{17} - 10^{19}$~eV. According to Fig.\ref{fig:4} the QGSJETII-04 model agrees better with the worldwide experimental data than QGSJET01. Estimations of the CR mass composition provided within the framework of this model reveal its rapid change towards lighter nuclei in the energy range $(1-20) \times 10^{17}$~eV. This probably is due to a transition from galactic to extragalactic component of UHECR. At $E_{0} \ge 2 \times 10^{18}$~eV our data and results displayed on Fig.\ref{fig:4}~\cite{Ulrich2001, Tunka2011, TA2013, HiRes2010, PAO2011, PAO2010, TA2011, TA2013} within measurement errors consistently indicate the slight change towards heavier nuclei of the mass composition with increase of primary energy.

\acknowledgements
The work has been conducted with the financial support from Russian Academy of Science within the framework of research program ``Fundamental properties of matter and Astrophysics'' and is supported by RFBR grant \#\,13--02--12036 ofi-m-2013.

\bibliographystyle{aipnum4-1}
\bibliography{masscomp-alter}

\end{document}